# Anisotropic electronic phase transition in CrN epitaxial thin films

Qiao Jin,[1,2] Jiali Zhao,[1] Manuel Roldan,[3] Shan Lin,[1,2] Shengru Chen,[1,2] Haitao Hong,[1,2] Yiyan Fan,[1] Dongke Rong,[1] Haizhong Guo,[4] Chen Ge,[1] Can Wang,[1,5] Jia-Ou Wang,[6] Shanmin Wang,[7] Kui-juan Jin,[1,2,5,*] and Er-Jia Guo[1,2,5,*]

[1] Beijing National Laboratory for Condensed Matter Physics and Institute of Physics, Chinese Academy of Sciences, Beijing 100190, China

[2] School of Physical Sciences, University of Chinese Academy of Sciences, Beijing 100190, China

[3] John M. Cowley Center for High Resolution Electron Microscopy, Arizona State University, AZ 85287, USA

[4] School of Physics and Microelectronics, Zhengzhou University, Zhengzhou 450001, China

[5] Songshan Lake Materials Laboratory, Dongguan, Guangdong 523808, China

[6] Institute of High Energy Physics, Chinese Academy of Sciences, Beijing 100049, China

[7] Department of Physics, Southern University of Science and Technology, Shenzhen, Guangdong 518055, China

*Correspondence and requests for materials should be addressed to K.J. and E.J.G. (Emails: kjjin@iphy.ac.cn and ejguo@iphy.ac.cn)

**Abstract:** Electronic phase transition in strongly correlated materials is extremely sensitive to the dimensionality and crystallographic orientations. Transition metal nitrides (TMNs) are seldom investigated due to the difficulty in fabricating the high-quality and stoichiometric single crystals. In this letter, we report the epitaxial growth and electronic properties of CrN films on different-oriented $NdGaO_3$ (NGO) substrates. Astonishingly, the CrN films grown on (110)-oriented NGO substrates maintain a metallic phase, whereas the CrN films grown on (010)-oriented NGO substrates are semiconducting. We attribute the unconventional electronic transition in the CrN films to the strongly correlation with epitaxial strain. The effective modulation of bandgap by the anisotropic strain triggers the metal-to-insulator transition consequently. This work provides a convenient approach to modify the electronic ground states of functional materials using anisotropic strain and further stimulates the investigations of TMNs.

**Main text**

Transition metal nitrides (TMNs) exhibit wide applications in the energy and coating fields due to their outstanding thermoelectricity,[1] catalysis,[2, 3] and mechanical properties.[4] However, the intrinsic physical properties of TMNs have not been well explored because the fabrication of stoichiometric single-crystalline TMNs is very challenge.[5, 6] Among the functional TMNs, chromium nitride (CrN) is an excellent metallic antiferromagnets and is readily applied in the next-generation spintronic and terahertz devices.[7] Bulk CrN has a paramagnetic cubic structure above near room temperature $T_N$ (~ 280 K) and transforms to antiferromagnetic orthorhombic structure below $T_N$.[8, 9] It serves as an alternative option to the well-established antiferromagnetic metallic alloys, such as $Mn_3Pt$[10] and $Mn_2Au$,[11, 12] because of its low price and robust in the ambient conditions. Therefore, the thoughtful understanding the magnetic and electrical properties of CrN is highly encouraging.

So far, the investigations of the transport behaviors of the CrN bulk and thin films are contradiction in the literatures.[1, 9, 13-18] Some groups report that CrN maintains semiconducting down to 5 K, while others demonstrate that it undergoes a insulator-to-semiconductor transition or insulator-to-metal transition with a sharp resistivity drop across $T_N$. These controversial results can be attributed to the crystalline impurities or nitrogen vacancies. Experimentally, it is challenging to fabricate high-quality stoichiometric CrN thin films by magnetron sputtering or laser splatting. The high-energy excited transition metal ions and high-vacuum environment unavoidably introduce the nitrogen deficiency in the as-grown CrN films. Recently, we developed a methodology of TMNs thin films growth using pulsed laser deposition technique assisted with an atomic nitrogen plasma source.[19] The nitrogen vacancies can be largely suppressed during the film growth. Intrinsic electrical properties can be investigated as a function of substrate-induced misfit strain and layer thickness. The conductivity in CrN layer maintains when its thickness reduces to one unit cell thick.[19]

Different from the biaxial strain induced by cubic substrates, the single layers grown on orthorhombic $NdGaO_3$ (NGO) substrates with various orientations suffer remarkably different in-plane strain. Therefore, the on-site Coulomb electronic correlation and band structure are also affected by the lattice distortion depending on the crystallographic orientations. In this letter, epitaxial CrN films are coherently grown on NGO substrates with different orientations. The transport properties and electronic states of CrN thin films exhibit remarkable different behaviors. Using X-ray linear dichroism (XLD) measurements, we attribute the anisotropic phase transition in CrN films to the strain-mediated electron redistribution between $e_g$ and $t_{2g}$ orbitals.

The CrN thin films with a thickness of ~25 nm were deposited on the NGO substrates with (001)-, (010)-, and (110)-orientations. To avoid the nitrogen vacancies in the CrN films, we used a stoichiometric ceramic CrN target for the laser ablation. The ceramic CrN was synthesized using a high-pressure reaction route. The heating process is carried out at 5 GPa and 1200 °C for 5 minutes. The powder was then sintered into a target with a diameter of 1 inch at 5 GPa and 1100 °C for 50 minutes.[20] Previously, we had determined the valence state of Cr ions to be +3 in CrN using X-ray absorption and X-ray photoemission spectroscopy. No apparent nitrogen vacancy was observed in the as-grown CrN films. The metallic phase maintains down to 5 K and phase transition occurs around ~ 280 K indicate the CrN films are clearly stoichiometric. We keep the optimal growth conditions used in our previous work.[19, 21] The growth temperature is 600 °C and the laser frequency and energy density are 5 Hz and ~1.5 J/cm$^2$, respectively. The CrN films were fabricated in vacuum to avoid the oxidization. The atomic nitrogen plasma source was used to compromise the nitrogen vacancies during the deposition. The growth rate was calculated by X-ray reflectivity fittings, meanwhile the film thickness was accurately controlled by counting the number of the laser pulses.

The structural characterizations of CrN films were performed on a Bruker D8 diffractometer with Cu target ($\lambda$ = 1.5406 Å). Figure 1(a) shows the X-ray diffraction (XRD) $\theta$-$2\theta$ scans of the CrN films grown on the NGO substrates with different orientations. Clearly, all the CrN films show single phase. The CrN films grown on (001)- and (110)-oriented NGO substrates exhibit (001) orientation, whereas the CrN films on (010)-oriented NGO substrates are (111)-oriented. The striking results are attributed to the minimization of misfit strain between CrN and NGO. CrN has a cubic rock-salt structure which its lattice parameter is $a$ = 4.15 Å, while NGO has an orthorhombic structure with *pbnm* space group which its lattice parameters are $a$ = 5.47 Å, $b$ = 5.50 Å, and $c$ = 7.71 Å. When CrN arrange along the (001) orientation, the in-plane lattice parameters perfectly match that of pseudocubic NGO (110) ($a_{(110)pc}$ ~ 3.868 Å), and that of pseudocubic NGO (001) ($a_{(001)pc}$ ~ 3.848 Å). The lattice mismatches between CrN and NGO substrates are -6.7% and -7.0% for (110)- and (001)-oriented NGO, respectively. However, when viewed along the [111] direction, CrN presents a hexagonal structure that $Cr^{3+}$ ions located in the vertex, which is closed to that of $Nd^{3+}$ and $Ga^{3+}$ enclosed in (100)- and (010)-oriented NGO. The distances between two nearby transition metal ions are 2.93 and 2.715 Å in the (111)-oriented CrN and pseudocubic (010)-oriented NGO, respectively. Therefore, the CrN films grown on (010)-oriented NGO substrates will preferentially arrange along the (111)-orientation, yielding to an extremely large compressive strain up to -7.3%. Figures 1(b) and 1(c) show the high-angle annular dark-field (HAADF) scanning transmission electron microscopy (STEM) images for the CrN films grown on the (010)- and (110)-oriented NGO substrates, respectively. Both samples exhibit the coherent growth on the NGO substrates with atomically sharp interfaces. The CrN films grown on the (110)-oriented NGO substrates show a cube-on-cube structure (Figure 1c), confirming its growth direction along the (001) orientation. While the CrN films grown on the (010)-oriented NGO substrates, the CrN films exhibit the layered structure by forming horizontal structural domains.[21] This unique structure is consistent with the (111)-oriented CrN films grown on the $Al_2O_3$ substrates.[21]

To explore the transport properties of the CrN films, we measured the temperature dependent resistivity ($\rho$) using standard van der Pauw method. All measurements were performed by a physical property measurement system (PPMS). Figure 2(a) shows the $\rho$-$T$ curves for the CrN films with various orientations. All the CrN films show the low resistivity ($\rho$ = 0.82-2.26 mΩ·cm) at room temperature, which are comparable to the results from other reports.[1, 13, 17] Strikingly, we observe a sharp difference in the transport behaviors of the CrN films with different orientations. For the (001)-oriented

CrN films, the metallic states appear at room temperature. As the temperature decreasing, they undergo a metal-to-insulator transition (MIT). $T_{MIT}$ for the CrN films grown on the (110)- and (001)-oriented NGO substrates are 70 and 150 K, respectively. However, the insulating phase in the CrN films grown on the (010)-oriented NGO substrates maintains throughout the temperature range measured. We find that the $\rho$-$T$ curves of all films cannot be described using a single activated model [$\ln(\rho) \sim 1/T$] or a single Mott's variable range hopping (VRH) model [$\ln(\rho) \sim (1/T)^n$]. [17] Previously, Catalan et al. [22] combine the normal activated conduction and 3D Mott's VRH to explain the complex transport behaviors,

$$\rho(T) = \rho_0 + A \exp\left(\frac{\Delta E}{k_B T}\right) + B \exp\left[\left(\frac{T_0}{T}\right)^{1/4}\right] \quad (1)$$

Where $\rho_0$, $A$, and $B$ are constants, $T_0$ is the temperature normalization constant. The second term describes the thermal excitation: $\rho(T) \sim \exp\left(\frac{\Delta E}{k_B T}\right)$, where $k_B$ is Boltzmann constant, and $\Delta E$ is the activation energy. The third term dominates the VRH: $\rho(T) \sim \exp\left[\left(\frac{T_0}{T}\right)^m\right]$, where the index $m = 1/4$ or $m = 1/3$ for three- or two-dimensional VRH. Figures 2(b) and 2(c) show excellent fits for the CrN films grown on the (110)- and (001)-oriented NGO substrates. We obtain $\Delta E \sim 49.5$ and $\sim 52.6$ meV for the CrN films grown on the (110)- and (001)-oriented NGO substrates, respectively. The increment of bandgap with the compressive strain also consistent with our earlier works. [15, 17] As shown in Figure 2(d), the $\rho$-$T$ curve of the (111)-oriented CrN films follows Eq. (1) below 125 K (solid black line), however, it transits to 2D Mott's VRH above 125 K (solid white line). Simultaneously, we find $\Delta E$ ($\sim 55.2$ meV) of the (111)-oriented CrN films is larger than those of the (001)-oriented CrN films.

Meanwhile, the room-temperature Hall mobility ($\mu$) and carrier density ($n$) of the CrN films are obtained. The CrN films exhibit a comparable $n \sim 1 \times 10^{20}$ cm$^{-3}$. The CrN films grown on the (110)-oriented NGO substrates show the maximum $\mu$ of 59.3 cm$^2 \cdot$V$^{-1} \cdot$s$^{-1}$, and the CrN films grown on the (010)-oriented NGO substrates have the lowest $\mu$ (48.2 cm$^2 \cdot$V$^{-1} \cdot$s$^{-1}$). These results suggest that strain-mediated lattice distortion affects the electron scattering in the CrN films even though the number of itinerant electrons keeps almost constant independent of crystallographic orientations.

To explore the intrinsic strain effect on the electronic states, we performed the X-ray absorption spectroscopy (XAS) and X-ray linear dichroism (XLD) measurements on the CrN films at the N *K*- and Cr *L*-edges at the room temperature. As shown in Figure 3(a), the centroid of the Cr $L_3$ peak for all the CrN films is $\sim$ 577.0 eV, the lineshape and peak positions of the N *K*-edges are consistent with those of other epitaxial CrN films. [15, 19] These results suggest that the valence state of Cr ions is 3+ and the CrN films are stoichiometric with negligible nitrogen vacancies. The XLD measurements were performed using X-ray beam aligned with angles ($\alpha$) of 90° and 30° respect to the sample's surface (Figure 3b). When $\alpha$ = 90°, i. e. the polarization of X-ray is parallel to the in-plane direction, the XAS reflects the $d_{x^2-y^2}$ orbital occupancy ($I_{\text{ip}} = I_{90°}$). When the $\alpha$ = 30°, XAS probes the unoccupied states in both $d_{x^2-y^2}$ and $d_{3z^2-r^2}$ orbitals, with $I_{\text{oop}} = [I_{90°} - I_{30°} \cdot \sin^2 30°)/\cos^2 30°]$. For simplifying the calculations, the signal of ($I_{90°} - I_{30°}$) can indirectly reflect the orbital asymmetry of $e_{\text{g}}$ band and is roughly proportional to the exact XLD signals. All the CrN films exhibit the positive XLD values, suggesting a higher occupancy in the $d_{3z^2-r^2}$ orbitals compared to that of the $d_{x^2-y^2}$ orbitals. The peak values of XLD scale with the compressive strain. The inset of Figure 3(b) shows the schematic of Cr: $d$ orbitals in CrN. The degenerated orbitals split into the upper two-fold $e_g$ bands ($d_{x^2-y^2}$ and $d_{3z^2-r^2}$) and lower three-fold $t_{2g}$ bands ($d_{xz}$, $d_{yz}$ and $d_{xy}$). Three electrons fill into the $t_{2g}$ bands with one electron filled into the $d_{xz}$, $d_{yz}$ and $d_{xy}$ orbitals and leave the upper-energy $d_{x^2-y^2}$ and $d_{3z^2-r^2}$ orbitals empty.[23] The compressive strain will reduce the atomic distance ($r$) between Cr ions, leading to the increase of $\Delta_{CF}$ accordingly. As increasing the compressive strain, the occupancy of itinerant electrons reduces, thus the conductivity of CrN films decreases. The localization of free carriers triggers the phase transition in the CrN films grown on (010)-oriented NGO substrates. A strain-mediated electronic anisotropy was also revealed in the SrFeO$_2$ films grown on the different substrates with the different strain states attributed to electron redistribution within degenerated orbitals in our previous work.[24]

In summary, we fabricated a series of high-quality CrN thin films on the different-oriented NGO substrates. The crystallographic arrangement and transport behaviors of the CrN films strongly depend on the substrates' orientations. The (001)-oriented CrN films exhibit a MIT as decreasing the temperature, whereas the (111)-oriented CrN maintains its insulating phase. We attribute the intriguing phase transition in the CrN films to the strain mediated electronic band structures. The compressive strain increases the bandgap and enhances the itinerant electron scattering, leading to the CrN transforms from 3D metallic state to 2D insulating state. Our results strengthen the important role of the epitaxial strain in the intrinsic physical properties in TMNs and unquestionably stimulate the investigation of TMNs towards electronic devices.

## Acknowledgements

This work was supported by the National Key Basic Research Program of China (Grant Nos. 2019YFA0308500 and 2020YFA0309100), the National Natural Science Foundation of China (Grant Nos. 11974390, 52025025, and 52072400), the Beijing Nova Program of Science and Technology (Grant No. Z191100001119112), the Beijing Natural Science Foundation (Grant No. 2202060), and the Strategic Priority Research Program (B) of the Chinese Academy of Sciences (Grant No. XDB33030200). The XAS experiments at the beam line 4B9B of the Beijing Synchrotron Radiation Facility (BSRF) of the Institute of High Energy Physics, Chinese Academy of Sciences were conducted via a user proposal.

**Figure and captions**

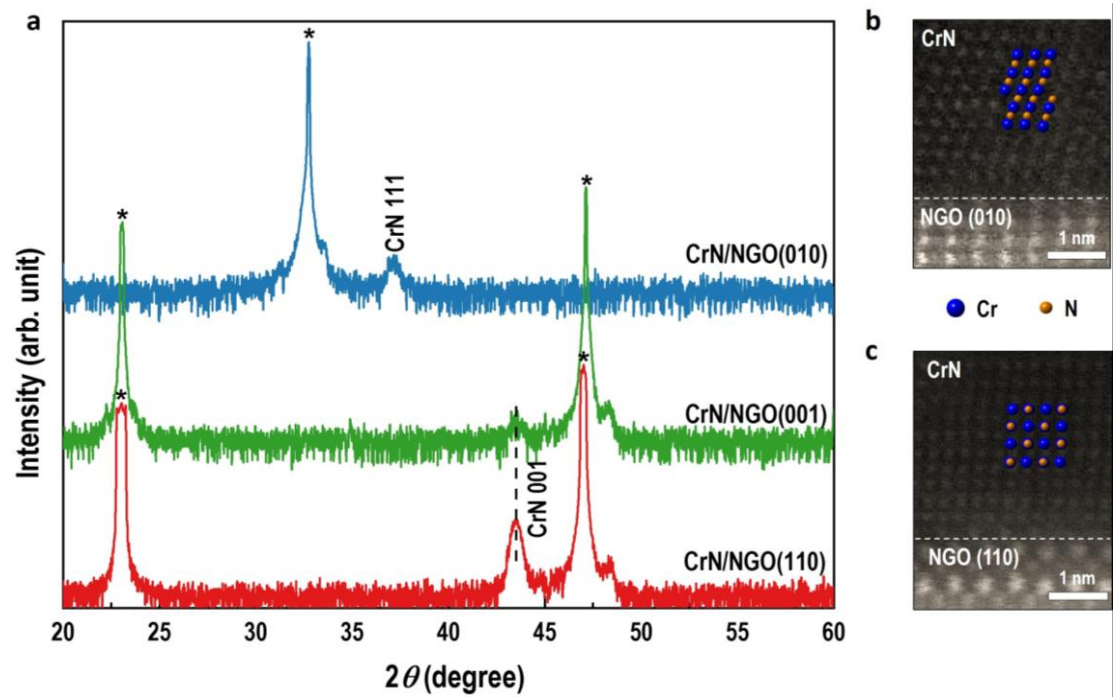

**Figure 1. Structural characterizations of CrN films grown on the different-oriented NGO substrates.** (a) XRD $\theta$-$2\theta$ scans of 25-nm-thick CrN films grown on (010)-, (001)-, and (110)-oriented NGO substrates. The asterisk symbols (*) indicate the diffraction peaks of substrates. (b) and (c) STEM images of CrN films grown on (010)- and (110)-oriented NGO substrates, respectively. The dashed lines indicate the abrupt interfaces between CrN and NGO.

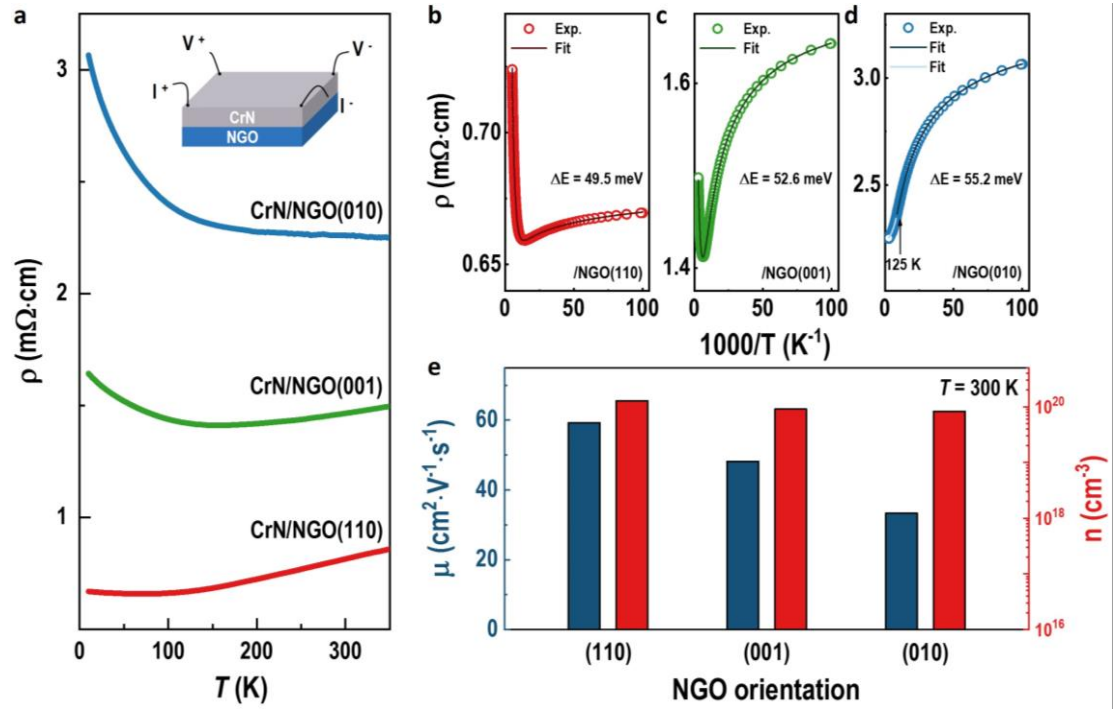

**Figure 2. Transport properties of CrN films.** (a) $\rho$-T curves of CrN films grown on (010)-, (001)-, and (110)-oriented NGO substrates. Inset: schematic of electrical measurements using conventional van der Pauw method. (b)-(d) Fitted curves of $\rho$-(1000/$T$) for CrN films grown on (110)-, (001)- and (010)-oriented NGO, respectively. (e) Room-temperature carrier mobility ($\mu$) and density ($n$) for CrN films.

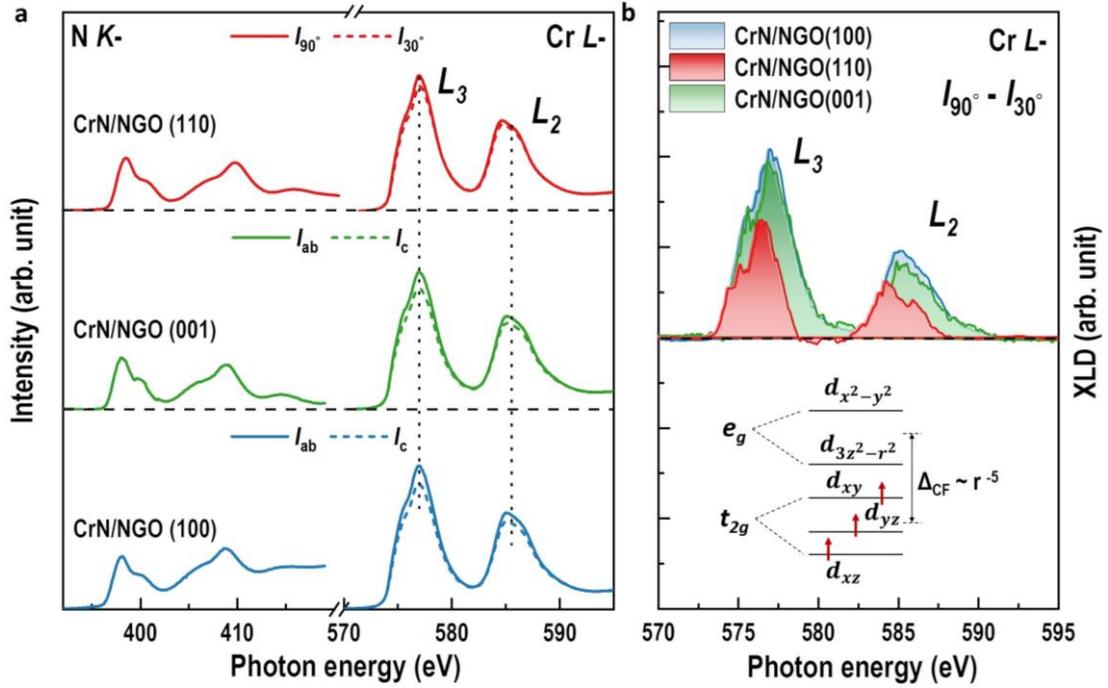

**Figure 3. Electronic state of CrN films.** (a) XAS and (b) XLD at N $K$- and Cr $L$-edges for CrN films grown on (110)-, (001)- and (010)-oriented NGO substrates. XAS were collected with x-ray beam aligned with angles ($\alpha$) of 90° and 30° respect to the sample's surface. When the $\alpha$ = 90°, i.e. the polarization of x-ray is parallel to the in-plane direction, the XAS reflects the $d_{x^2-y^2}$ orbital occupancy ($I_{ip} = I_{90°}$). When the $\alpha$ = 30°, XAS probes the unoccupied states in both $d_{x^2-y^2}$ and $d_{3z^2-r^2}$ orbitals, with $I_{oop} = [I_{90°} - I_{30°} \cdot \sin^2 30°]/\cos^2 30°$]. For simplifying the calculations, the signal of ($I_{90°} - I_{30°}$) can indirectly reflect the orbital asymmetry of $e_g$ band and is roughly proportional to the exact XLD signals. Inset: schematic diagram of Cr: $d$ orbitals in CrN. The crystal field splitting energy ($\Delta_{CF}$) is inversely proportional to the distance ($r^5$) between two Cr atoms.

15